\documentclass{nature}
\usepackage[ngerman,british]{babel}
\usepackage{amsmath}
\usepackage{graphicx}
\usepackage{bm}
\usepackage{hyperref}
\usepackage[mathlines]{lineno}
\usepackage{tikz}
\usepackage{caption,setspace}
\usepackage{verbatim}

\title{\textbf{The universality in urban commuting across and within cities}}

\author{Lei Dong$^{1,2,3}$, Paolo Santi$^{2,4\ast}$, Yu Liu$^{1\ast}$, Siqi Zheng$^{3}$, Carlo Ratti$^{2}$}

\begin{document}
\maketitle

\begin{affiliations}
\item{\small Institute of Remote Sensing and Geographical Information Systems, School of Earth and Space Sciences, Peking University, Beijing 100871, China}
\item{\small Senseable City Laboratory, Department of Urban Studies and Planning, Massachusetts Institute of Technology, Cambridge, MA 02139, USA}
\item{\small Sustainable Urbanization Laboratory, Department of Urban Studies and Planning, Massachusetts Institute of Technology, Cambridge, MA 02139, USA}
\item{\small Istituto di Informatica e Telematica del CNR, Pisa 56124, Italy}
\item[$^*$]{\small Corresponding authors: P.S. (psanti@mit.edu) and Y.L. (liuyu@urban.pku.edu.cn).}
\end{affiliations}

\setlength{\parskip}{1.2em}

\begin{abstract}
Commuting is a key mechanism that governs the dynamics of cities. Despite its importance, very little is known of the properties and mechanisms underlying this crucial urban process. Here, we capitalize on $\sim$ 50 million individuals' smartphone data from 234 Chinese cities to show that urban commuting obeys remarkable regularities. These regularities can be generalized as two laws: (i) the scale-invariance of the average commuting distance across cities, which is a long-awaited validation of Marchetti's constant conjecture, and (ii) a universal inverted U-shape of the commuting distance as a function of the distance from the city centre within cities, indicating that the city centre's attraction is bounded. Motivated by such empirical findings, we develop a simple urban growth model that connects individual-level mobility choices with macroscopic urban spatial structure and faithfully explains both commuting laws. Our results further show that the scale-invariants of human mobility will ultimately lead to the polycentric transition in cities, which could be used to better inform urban development strategies.
\end{abstract}

\section*{Introduction}
\noindent Urban commuting, i.e., the daily mobility behaviour of individuals who travel between their residential area and working area, is an important mechanism that governs the dynamics of cities\cite{batty2013new,mazzoli2019field}, constituting the `flowing’ backbone of a city\cite{berry1969metropolitan,anas1998urban,louail2015uncovering}. Commuting is also a major contributor to traffic congestion\cite{barthelemy2016structure}, and is an accelerant of the spread of infectious diseases within cities\cite{tizzoni2014use,chang2020mobility}. Despite the crucial role of urban commuting, our understanding of human commuting is surprisingly limited, and many puzzles about commuting remain unsolved. One key puzzle is that, despite the dramatic increase in the size of many cities in terms of population and area, the commuting distance and time appear to have remained remarkably stable\cite{gordon1991commuting,levinson1994rational,anas2015urban,kung2014exploring,louf2014congestion}, as suggested by Marchetti's constant hypothesis\cite{tanner1961factors,szalai1972use,zahavi1979umot,marchetti1994anthropological}. The conserved commuting properties are quite counter-intuitive, as the distance from the periphery to the urban centre in larger cities is obviously greater than that in medium- or small-sized cities.

In recent decades, traffic survey data from some developed countries have shown evidence of conserved mobility behaviour\cite{redding2015transportation}. For instance, the average travel time in the UK has been remarkably steady for more than 30 years\cite{metz2008limits}. The American Community Survey also demonstrates that the commute time in the US has remained relatively consistent\cite{commuting}. Further, a recent study notes that the typical number of locations visited by an individual within a given time window is scale-invariant across cities of different sizes\cite{alessandretti2018evidence} (see Supplementary Information section I for a detailed review of the empirical findings in previous literature). However, previous empirical evidence is qualitative and inconclusive, and it has been collected for single cities or a small set of cities\cite{levinson1994rational,anas2015urban,alessandretti2018evidence,huang2018tracking}. Because of the difficulty of collecting granular mobility data at a large-scale (e.g., nation-wide), little is known about how human mobility varies across and within cities that differ vastly in size. Furthermore, the mechanisms underlying the observed scale-invariant behaviour and their connection to the spatial structure of cities are still not fully understood. 

With the development of communications technology, especially the large-scale use of smartphones, we can now obtain real-time, high-resolution mobility data that enable us to decode urban commuting behaviour systematically\cite{kraemer2020mapping}. Here, by analysing approximately 50 million individuals' mobile phone location data, we show that urban commuting obeys remarkable regularities that apply to highly diverse cities (see Extended Data Fig. 1 and Methods). These regularities can be summarized as two laws: (i) commuting distance at the city level is independent of city size (the scale-invariance law); and (ii) commuting distance within a city follows an inverted U-shaped function of the distance from the city centre, which implies that the city centre has an upper bound on the employment attractiveness of residents (the attractiveness law). We propose an urban growth model that shows that the urban commuting laws accord well with the cost-benefit trade-off of an individual's location choice, establishing a link between individual-level commuting choices and the resulting spatial structure at the city level.

\section*{Results}

\subsection{Scale-invariance of commuting distance}\

\noindent We first show that the most straightforward quantity, the average commuting distance, $\langle C_d \rangle$, is scale-invariant across cities (we consider prefecture-level cities in this study). To demonstrate this result, we detect individuals' home and work locations by means of a supervised machine learning algorithm and use the geodesic distance between the home and work coordinates as a measure of the commuting distance (see `Home/work location detection' in Methods and Extended Data Figs. 2, 3). Notably, the geodesic distance is a simplification of reality, that neglects some minor details, such as the spatial organization of the road network, transportation mode choice of each trip, and traffic conditions; however, it is sufficient to derive useful information about the spatial pattern of a city (Fig.~\ref{fig:1}a). Additonally, geodesic distance is the most commonly used parameter in the literature to measure commuting and urban spatial structures \cite{anas1998urban,glaeser2008cities}. 

One major concern in using mobile phone data to analyse urban commuting is the relevance of mobile data to \textit{the real data}. To address this issue, we validate mobile phone data by comparing them to three datasets -- individual self-reported home/work locations, the resident population obtained from micro-census at the district level, and a transportation survey of Shanghai (see `Data validation' in Methods for details). These systematic validations show that the dataset and algorithm in this paper are among the most comprehensive micro-data for urban commuting currently available.

As Fig.~\ref{fig:1} shows, although city size varies over two orders of magnitude, the average commuting distance $\langle C_d \rangle$ is remarkably similar across cities in a statistical sense. Quantitatively, $\langle C_d \rangle$ is 7.84 (0.0663) km, and the correlation between city population and $\langle C_d \rangle$ is not significant (Fig.~\ref{fig:1}b, $p$-value = 0.327, $r^2 < 0.01$; see Extended Data Fig. 4 for a robustness check). This result is further confirmed by replacing population size with built-up area, another important indicator of city size (Fig.~\ref{fig:1}c, $p$-value = 0.916, $r^2 < 0.01$). This scale-invariant finding is surprising, as we might expect larger cities to have longer commuting distances. However, even though Shanghai's population is ten times larger than that of Daqing (an industrial city in northeastern China), their average commuting distances are very similar (8.82 km in Shanghai and 8.64 km in Daqing). Notably, we cannot calculate specific commuting times for each city because of the lack of information on individual transportation modes (e.g., walking, car, bus, and subway). Considering that congestion is more severe in large cities\cite{louf2014congestion}, it is possible that commuting time is greater in large cities than in small and medium cities for a given commuting distance. 

In addition to commuting distance, we find that the average number of locations visited by residents in a city (denoted as $\langle S \rangle$) is stable and independent of population size (Fig.~\ref{fig:1}d, $r^2 < 0.01$, $p$-value = 0.688) or built-up area (Fig.~\ref{fig:1}e, $r^2 < 0.01$, $p$-value = 0.264), even though large cities have more {\it explorable spaces}. Here, a visit is defined as staying within a certain spatial range (200 m) for a certain period (10 min); see Methods for details.

\subsection{The upper bound of the city centre's attraction}\

\noindent The invariant values $\langle C_d \rangle$ and $\langle S \rangle$ observed here are the means of a set of distributions, which are similar in different cities. Within a city, commuting behaviour does vary across locations. Remarkably, within a city, the spatial distribution of commuting also obeys a universal pattern: from the centre to the suburbs, the intra-city commuting distance $\langle d_c \rangle$ and time $\langle d_t \rangle$ show an inverted U-shaped curve, indicating that the city centre exerts a bounded employment attraction on urban dwellers. People outside this attraction bound are most likely to find a job in a closer sub-centre than to travel to the main centre -- in accordance with known spatial economics theory\cite{fujita2001spatial}.

To investigate the intra-city commuting pattern, we partition the space into 1 km $\times$ 1 km grid cells and compute the commuting matrix (also called the origin-destination matrix) between cells. By querying online map services, we calculate the real-time commuting distance and time by public transit and driving (see Methods).  Figure~\ref{fig:2} shows the spatial distributions of commuting distance for Shanghai and Chengdu, and Extended Data Fig. 5 and Extended Data Fig. 6 present the same results for additional cities (Extended Data Table 1 presents descriptive statistics for these cities). These figures demonstrate that the average commuting distance of people living in the city centre is relatively short. As the distance from the place of residence to the city centre increases, the commuting distance increases. Nevertheless, this increase is not unlimited: beyond a certain distance from the city centre, the commuting distance decreases again. 

To generalize these observed commuting patterns, we calculate the relationship between the distance from the city centre and the average commute distance and time, $\langle c_d \rangle$ and $\langle c_t \rangle$, for a corresponding distance. Figure~\ref{fig:2} shows the average commuting distance (Fig.~\ref{fig:2}b, e) and time (Fig.~\ref{fig:2}c, f) by public transit and driving, respectively. The results provide a clear finding: there is an inverted U-shape curve between distance from city centre and commuting costs. The form of the curve indicates the existence of an upper bound on commuting costs, and it implies an upper bound on the city centre's employment {\it attraction}, measured by the peaked distance from the city centre. Specifically, for Shanghai, this upper limit is reached at a radius of $\sim 12$ kilometres from the city centre. For Chengdu, this radius is $\sim 10$ kilometres. The analysis of dozens of cities indicates that the upper limit of attraction is approximately 10-15 kilometres, the upper bound of commuting distance is approximately 10 kilometres, and the upper bound of commuting time is 50-60 minutes by public transit or 30-40 minutes by driving (see Extended Data Fig. 5 and Extended Data Fig. 6). Note that because we do not know the transportation mode of each individual, we have calculated the intra-city commute using driving and public transportation separately. The similar shape of the calculated curves (i.e., inverted U-shape) is sufficient reason to believe that commuting within cities -- often a mixture of these two modes -- follows the same pattern. 

The existence of these upper bounds does not imply that a city's outward expansion will stop after reaching a specified distance from the centre. Instead, new development will spill into sub-centres when the commute to the main centre from fringe areas exceeds certain constraints. In other words, cities are `forced' to switch from a monocentric to a polycentric model, and such a process is driven by commuting constraints. In practice, however, this polycentric transition is driven by a combination of top-down government policy and bottom-up market forces\cite{batty2008size,bertaud2018order}.

\subsection{The urban growth model}\

\noindent The empirical results indicate that at the city level, commuting distance and average number of visited locations are not influenced by city size. Inside a city, we observe a bounded attraction of the city centre and consequent polycentric urban structure. To explain the observed individual commuting behaviour with a macro-level urban structure, we propose a spatial growth model. This model is similar to existing urban economics models\cite{fujita2001spatial} but also adopts the out-of-equilibrium characteristics of statistical physical models\cite{louf2013modeling,zhang2015scaling}. In our model, a city's population is an exogenous variable that is added to the system according to certain growth rules and is modelled as an indivisible population unit (i.e., household). When a new population unit joins the city, the home location is given, and a work location is chosen in consideration of the wages provided by potential workplaces and the commuting costs (Fig.~\ref{fig:3}). The model is detailed in the Supplementary Information; we present the main assumptions and results here.

\textbf{Matching growth.} 
The first node (population unit) is located at the centre of a two-dimensional space (i.e., a city). At each time step $t$, a new node $j$ is generated randomly. This node survives only if it matches with existing ones $i$; otherwise, it will be removed. Here, matching means the distance between nodes $i$ and $j$ is less than a radius $\ell$ (Fig.~\ref{fig:3}). Under this growth rule, the radius of the generated node cluster at time $t$ is proportional to the time step: $R(t) \sim t$, as $t \rightarrow  \infty$. The total number of population units in the system is $N(t) \sim R(t)^{3}$. Thus, we have the following relation between city size and total population: $R(t) \sim N(t)^{1/3}$ (see Supplementary Information). This matching growth mechanism was first proposed in ref.\cite{zhang2015scaling} to explain online community growth and was subsequently used to explain urban scaling laws\cite{li2017simple}. The mechanism captures both the densification and expansion during urban growth and is sufficiently simple to allow analytic solutions for the distribution of population sizes. 

\textbf{Constant connection.} 
Each node's home location is defined as the position where it was generated in the system. Each node is connected to $k$ existing locations (nodes), one of which is the working place (detailed below), and the remaining $k-1$ locations are randomly selected among neighbours within radius $\ell$. Infrastructure (road) within cities is built to support people's connections. The local road length density is $\delta(r, \theta, t)$, which varies with the square root of the local population density, $\delta(r, \theta, t) \sim \rho(r, \theta, t)^{1/2}$ (Extended Data Fig. 8). Local road volume is assumed to be the number of visits multiplied by the local road length density. Thus, the total road volume is $V(t) = \int_0^{R(t)} k \cdot 2\pi r\rho(r,\theta,t)^{1/2} dr \sim R(t)^{5/2}$. Recall that $R(t) \sim N(t)^{1/3}$, which implies a sublinear relation between population and road volume $V(t) \sim N(t)^{5/6}$.

\textbf{Utility function of workplace choice.} 
Following the setting of ref.\cite{louf2013modeling}, we assume that individual $j$'s utility function of workplace choice $U(Z)$ is a simplified version of the Fujita and Ogawa model\cite{fujita1982multiple}

\begin{equation}
Z(i,j) = W(i) - C_{T}(i,j),
\label{z1}
\end{equation}

\noindent where $W(i) = \alpha E(i)^{\gamma}$ is the average wage of location $i$, $E(i)$ is the employment population at location $i$, $\gamma$ is a scaling exponent that captures agglomeration and can be estimated by means of empirical data (we set $\gamma = 0.15$ in the simulation, see Supplementary Information and ref.\cite{prud1999size,bettencourt2013origins}), and $\alpha$ is a normalized parameter interpreted as the {\it base wage}. $C_{T}(i,j)$ is the commuting cost between $i$ and $j$, which is estimated by the Bureau of Public Roads (BPR) with a simple formula\cite{branston1976link}, $ C_T(i,j) = \beta \frac{d_{ij}}{v_{0}}[1+b(\frac{T_{ij}}{c})^\mu]$. In this formula, $T_{ij}$ is the traffic between $i$ and $j$, $v_0$ is the free-flow travel velocity, $c$ is the typical capacity of a road, $\mu$ quantifies the resilience of the transportation network to congestion, $b$ is a scale factor that is suggested to be 0.15 by BPR\cite{branston1976link}, and $\beta$ is a parameter quantifying \textit{the unit value of commuting time}. We further simplify the formula by assuming that traffic $T_{ij}$ is a function of only the centre $i$ and therefore rewrite $T_{ij} = T(i)$ as the total traffic incoming to centre $i$\cite{louf2013modeling}. $T(i)$ should be distributed across roads around centre $i$. We have calculated the scaling relation between road volume and population as $V(i) \sim T(i)^{1 - \eta}$, where $\eta = 1/6$. Thus, the road-average traffic around centre $i$ is $T(i)/V(i) \sim T(i)^{\eta}$. Eq. (\ref{z1}) can then be rewritten as

\begin{equation}
    Z_{ij}^{*} = E(i)^{\gamma} - \beta^{*} d_{ij}[1 + b(\frac{T(i)^{\eta}}{c})^{\mu}],
\label{z2}
\end{equation}

\noindent where $\beta^{*} = \beta / (\alpha v_0)$ measures the ratio of unit commuting cost $\beta$ to the base wage $\alpha$. Eq. (\ref{z2}) is the key equation of the model and depicts a cost-benefit trade-off, as shown in some spatial network models\cite{louf2013emergence}. When $\beta^{*} \rightarrow 0$, the commuting cost is negligible compared to the wage term, all individuals will commute to the city centre to work, and we obtain a monocentric pattern composed of one hub at the city centre (Fig.~\ref{fig:4}a). When $\beta^{*} \rightarrow +\infty $, the wage term is negligible compared to the commuting cost, individuals will work at the nearest node, and the city is decentralized (Fig.~\ref{fig:4}d). Importantly, between these two extremes, there is a critical value $\beta^*$ at which a polycentric phase will emerge (Fig.~\ref{fig:4}b). Figure~\ref{fig:4}d shows this phase transition by plotting the ratio of the largest employment centre and the second-largest employment centre (insert) with different values of $\beta^*$.

Figure~\ref{fig:4}e presents the relation between $\beta^*$ and the average commuting distance with different population sizes. At the monocentric stage, the average commuting distance increases with the population size (left side of Fig.~\ref{fig:4}e). However, the commuting distance becomes scale-invariant when the system steps into the polycentric structure - the curves of different population sizes collapse into a single curve (also shown in the inset of Fig.~\ref{fig:4}e), consistent with empirical findings in Fig.~\ref{fig:1}. The within-city heterogeneity of commuting flows is depicted in Fig.~\ref{fig:4}f. From the city centre to the suburbs, the commuting distance increases linearly in the monocentric phase (black), flattens in the decentralized phase (blue), and reaches a peak before subsequently declining in the polycentric phase (red and green) -- the inverse U-shape curve produced by the model is very similar to the empirical findings reported in Fig.~\ref{fig:2}. In the real world, $\beta^*$ can be estimated by the ratio of transportation expenditure to total income, and this ratio is approximately 0.15-0.2 in different countries according to a survey (Supplementary Table 3). This constant number echoes the findings of in our polycentric model (Fig.~\ref{fig:4}b, e).

Beyond numerical simulation, we can derive several important analytical results from this model, such as the population size when a city evolves into a polycentric stage and the attraction radius of the city centre (Methods). Notably, the attraction radius of the largest centre is derived as $d \sim (c^\mu/\beta^*)^{\frac{1/3}{\mu\eta - \gamma + 1/3}}$. This formula indicates that the most effective way to stretch the attraction radius is to increase the transportation capacity $c$ (as $\mu$, $\eta$, and $\gamma$ are approximately stable across cities), for example, by building more roads and adding a subway system to cities (as shown in Fig.~\ref{fig:1}b). The commuting curve shifts to the right if we increase $c$ (see the green curve in Fig.~\ref{fig:4}f). Meanwhile, if the agglomeration effect $\gamma$ of a city is large, the city should be expected to have a larger attraction radius.

\section*{Discussion}
Urban spatial structure is a crucial research topic in urban economics and urban planning, and many classical spatial structure models have been proposed (e.g., Alonso-Muth-Mills\cite{brueckner1987structure} model and Fujita-Ogawa\cite{fujita1982multiple} model). However, these classic urban models do not couple the dynamics of individual mobility behaviour, and they treat a city as being in an equilibrium characterised by static spatial distributions of residents and firms. Nonetheless, the equilibrium assumption does not hold, as cities are typically out-of-equilibrium systems. The model presented herein overcomes these shortcomings and introduces a growth mechanism that can simulate three spatial structure phases by adjusting a single parameter $\beta^*$. Our growth model also reproduces the quantitative results derived from empirical data, i.e., the constant commuting distance and the inverted-U shape of commuting curves. For simplicity, we make a number of strong assumptions. Therefore, further investigation can be conducted to extend the model by relaxing these assumptions. For example, instead of choosing the location with the highest utility when choosing a workplace, individuals can choose different locations with a probability that is proportional to the utility function, similar to the spatial preferential attachment mechanism\cite{balister2018topological}. Furthermore, the assumption that each individual has the same preferences can be removed by considering idiosyncratic utility shocks\cite{ahlfeldt2015economics}.


Given the complexities that influence human mobility – heterogeneous distributions of population and services, transportation, geography, etc. – one would not expect urban commuting to obey universal patterns that can be explained by a simple microscopic model. This universality is likely due to mobility constraints that are rooted in human biology\cite{kristal2020we}. Interestingly, similar constraints exist not only in mobility behaviour but also in psychology (e.g., Dunbar number\cite{dunbar1992neocortex}), social networks (e.g., six degrees of separation\cite{newman2011structure}), and economics (e.g., the natural rate of unemployment\cite{barnichon2017natural}). 

In summary, the approaches presented in this paper quantify the scale-invariant characteristics of human commuting on a comprehensive mobile phone dataset and model the connection between micro-level mobility behaviour and the macro-level spatial structure of a city. From a practical perspective, urban planners and administrators could use the results to better inform urban development strategies, as scale-invariants of human mobility will ultimately lead to a polycentric transition in cities. From a theoretical perspective, the spatial model proposed here could be extended to other fields where macro-level structures result from micro-level decisions, such as biology and ecology\cite{gisiger2001scale}.

\bibliography{ref}
\bibliographystyle{naturemag}

\section*{Acknowledgments}
We would like to thank Meng Li and Hongbin Pei for providing help in data preprocessing. We also would like to thank Gabriel Kreindler, Kevin O'Keeffe, Youguo Qin, Geoffrey West, Haishan Wu, Yunhan Yang and Jiang Zhang for helpful discussions; and all the members of the MIT Senseable City Lab Consortium and the Sustainable Urbanization Lab for supporting this research. L.D. and Y.Liu were supported by the National Natural Science Foundation of China (41625003, 41801299, 41830645).

\section*{Author contributions} All authors contributed to the design of the research and the writing of the manuscript. L.D. performed the research, analysed data and model, with P.S. supervision. 

\section*{Competing interests} The authors declare no competing interest.

\section*{Data and code availability}
The raw mobile phone datasets used in this research are not available for distribution due to non-disclosure agreements and privacy issues. The aggregated data and code, to replicated results in this article, are available through https://github.com/leiii/commute.

\clearpage
\begin{figure*}[!ht]
    \centering
    \includegraphics[width=.8\linewidth]{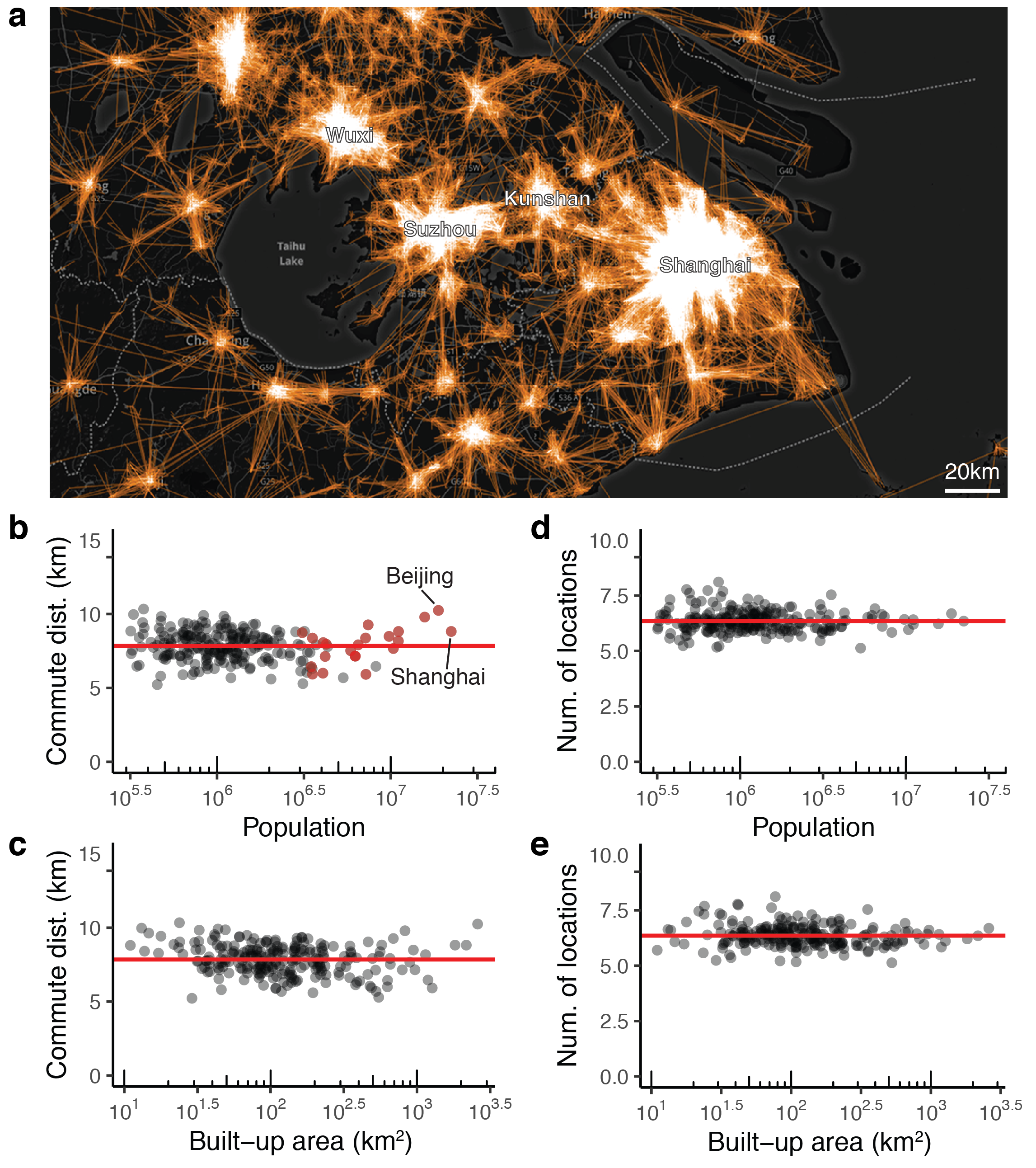}
    \caption{\textbf{Scale-invariance of urban commuting at the city level}. \textbf{a,} Spatial structure of commuting flows of the Yangtze River Delta. \textbf{b,} Correlation between commuting distance and population size. Each point represents one city. Statistical summary: n = 234, mean = 7.84, s.e. = 0.0663, $p$-value = 0.327, $r^2$ = 0.00414. Note that cities with subway systems are coloured in red. \textbf{c,} Similar to (\textbf{b}), the correlation between commuting distance and built-up area. Statistical summary: n = 234, $p$-value = 0.916, $r^2 < 0.01$. \textbf{d,} Average number of visited locations and population size. Each data point represents one city. Statistical summary: n = 234, $p$-value = 0.688, $r^2 < 0.001$. \textbf{e,} Similar to (\textbf{d}), the correlation between the average number of visited locations and built-up area (n = 234, $p$-value = 0.264). The city in which each user lives is defined as the city in which most user's stay points are located. The urban population of each city is based on the 2010 Census. A city's built-up area data are downloaded from the World Bank PUMA database.}
    \label{fig:1}
\end{figure*}

\begin{figure*}[!ht]
    \centering
    \includegraphics[width=1.\linewidth]{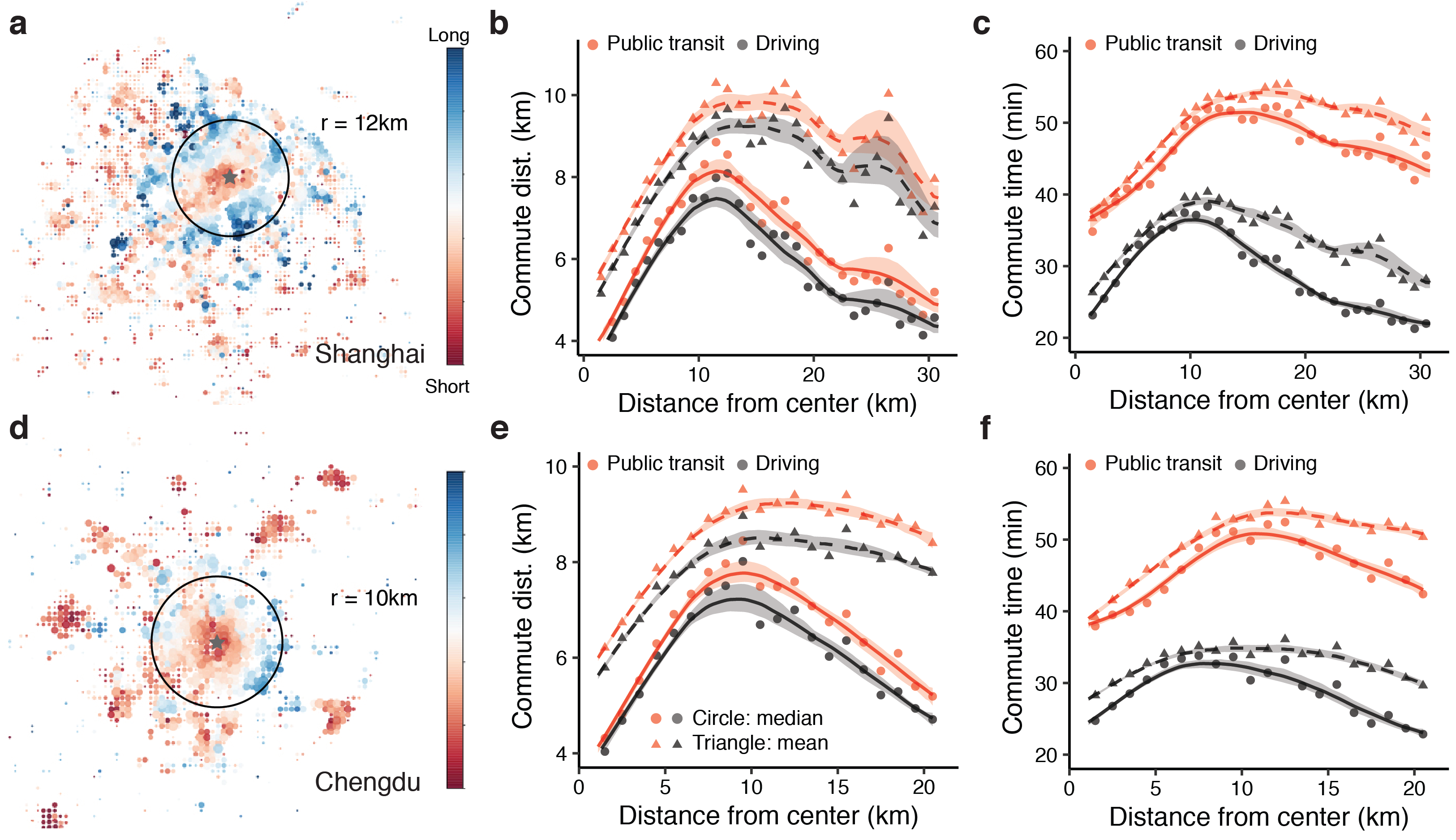}
    \caption{\textbf{Commuting within a city}. Spatial distribution (by home location) of commuting distance in Shanghai (\textbf{a}) and Chengdu (\textbf{d}). The city centre is marked with a star symbol. Average commuting distance/time (y) with varying distance from the city centre (x) of Shanghai (\textbf{b, c}) and Chengdu (\textbf{e, f}). The red colour refers to commuting by public transit, black by driving, and circle and triangle symbols represent the mean and median results, respectively. Note that we add 10 mins parking time to the driving mode, which does not change the shape of the curve but makes the commuting time more realistic. We also apply non-parametric fitting for each plot and calculate the 95\% confidence interval (shaded areas) of each fitting curve by the bootstrapping method in R\cite{sestelo2017npregfast}.}
    \label{fig:2}
\end{figure*}

\begin{figure*}[!ht]
    \centering
    \includegraphics[width=.6\linewidth]{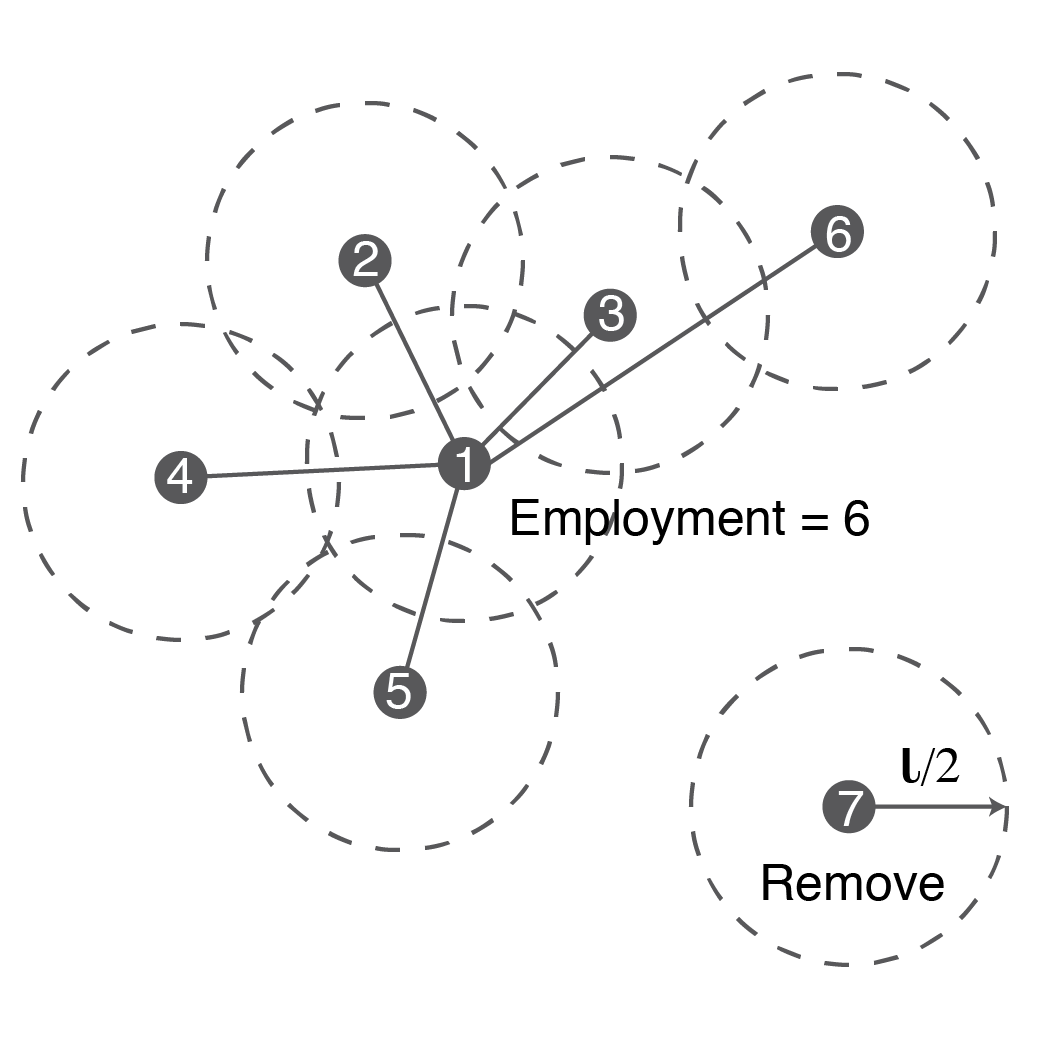}
    \caption{\textbf{Model schematic.} Only if a new node, representing a population unit, matches (the nearest distance between a new node and the existing ones is less than the radius $\ell$, buffered with a dash line) with the existing ones does it survive (e.g., Node 1-6); otherwise, it will be removed (e.g., Node 7). Then, the new node will choose a workplace according to the utility function detailed in the main text, generating a direct link to the work location.}
    \label{fig:3}
\end{figure*}

\begin{figure*}[!ht]
    \centering
    \includegraphics[width=1.\linewidth]{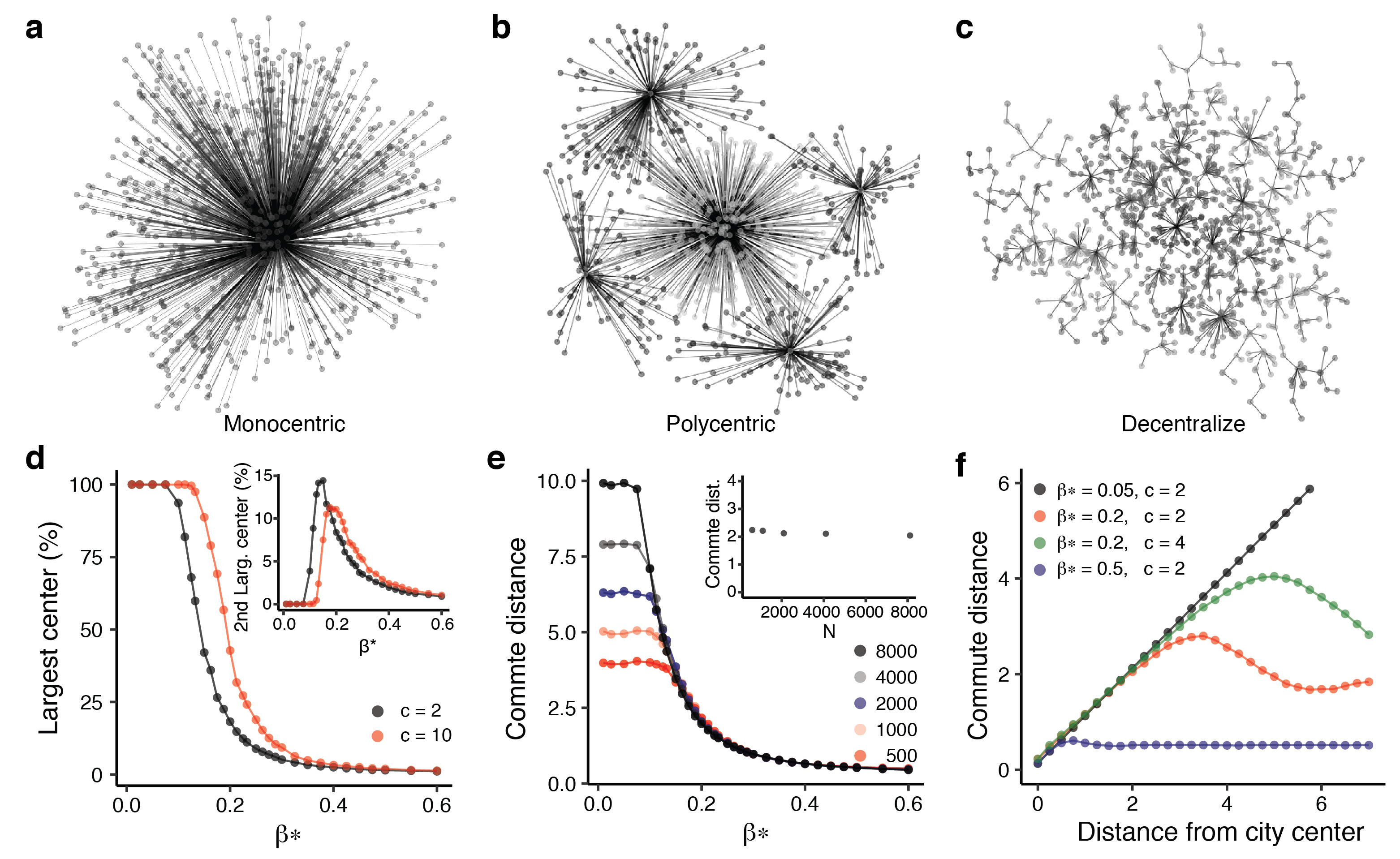}
    \caption{\textbf{Model and simulation.} \textbf{a-c,} Schematic of monocentric (\textbf{a}), polycentric (\textbf{b}), and decentralized (\textbf{c}) cities generated by simulation with $N=1000$, $\gamma=0.15$, $c=2$, $\mu=2$, and $\beta^* = 0.05,\ 0.2,\ 0.5$. \textbf{d,} The node percentage of the largest cluster and the second largest cluster (insert) and $\beta^*$ with $c=2,10$ and $\mu=2$ ($N=4000$). The peak of the second largest cluster indicates the phase transition of the system. \textbf{e,} Commuting distance and population size. Commuting distance is scale-invariant when the system has a the polycentric structure - the curves for different population sizes collapse into a single curve (also shown in the inset of (\textbf{e}), where $\beta^* = 0.2$). \textbf{f,} Inverted U-shaped curve of commuting distance.  From the city centre to the suburbs, the commuting distance increases linearly in the monocentric phase (black), flattens in the decentralized phase (blue), and reaches a peak then declines in the polycentric phase (red and green). The peak of the commuting curve shifts to relatively longer distances if road capacity $c$ is increased. The results in (\textbf{d-f}) are averaged over 50 simulations.}
    \label{fig:4}
\end{figure*}

\clearpage

\section*{Methods}

\subsection{Data description and pre-processing}\

\noindent The mobile phone dataset used here is provided by a large location-based service provider in China. The raw data are generated when people use location-based services (e.g., searching online maps, calling a ride-sharing car, or asking for delivery services). Each location point includes an anonymised user ID, longitude, latitude, and timestamp. Extended Data Fig. 1 shows that the number of mobile phone subscribers per city is highly correlated with the census data ($r = 0.903$), meaning that our dataset is well represented in terms of population coverage.

\subsubsection*{Home/work location detection}
To infer visited, home, and work locations for each mobile phone user, we adopt the following steps: 

\begin{itemize}
    \item \textbf{Stay point detection.} For each anonymous individual, we have a series of geo-positioning points with timestamps. A stay point is defined as a location at which the user moves less than $d = 200$ m in a minimum time period of $t = 10$ min. The average number of stay points for each user is 527 (s.d. 472). According to our previous study\cite{dong2017measuring}, the stay point detection algorithm is robust in adjusting these thresholds. 
    \item \textbf{Clustering and visitation.} We then cluster stay points into different clusters using the density-based spatial clustering of applications with noise (DBSCAN) clustering algorithm\cite{ester1996density} (the two parameters for DBSCAN are MinPoint = 1 and radius = 200 m). These clusters are defined as stay locations, which we use in Fig.~\ref{fig:1}d, e of the main text. Since mobile phone data are sparsely sampled in the temporal scale, the total number of visited locations in the interval, which comprises the entire observation period, is likely underestimated. However, the goal of our analysis is to test the scale-invariant hypothesis, and the results should not be affected by the sparsity of the dataset (the sampling methods are the same across all the cities).
    \item \textbf{Feature extraction.} We assume that home and work are among the stay locations. Thus, we extract 28 features from the individual-level data to the cluster-level data (Extended Data Table 1). To increase the prediction accuracy for home/work detection, we drop clusters with only one stay point (i.e. MinPoint = 2)\cite{dong2017measuring}. Users who have only one stay point are usually the result of sparse data or data generated by a fixed device (e.g., WiFi hotspots).
    \item \textbf{Classification.} Finally, we match the self-reported home and work data with the extracted features and split the dataset into a training set (80\%) and a testing set (20\%). We then apply Xgboost\cite{chen2016xgboost}, a tree-based supervised machine learning algorithm, to train two classifiers for home and work classification. Extended Data Fig. 2 presents the spatial distributions of detected home and work locations. Extended Data Fig. 3 maps the commuting networks based on these home and work locations. Note that we remove users whose home and work locations have a geodesic distance longer than 100 km, as such a long distance is uncommon in urban commuting (in our dataset, such an anomaly could be caused by users changing jobs or residences). After filtering, we have 32.3 million users remain.
\end{itemize}

\subsection{Data validation}\

Many previous studies have used a rule-based approach to identify home and work locations, for example, setting the location where the user most often stays at night/day as home/work. However, this approach lacks validation and is challenging to assess in terms of accuracy.

Since we have self-reported home and work addresses for approximately 10,000 users, we can use supervised machine learning methods to identify home and work and to evaluate the accuracy of the identification method more accurately. We calculated the accuracy on the test set according to the following formula:

\begin{equation}
    Accuracy = \frac{\# \ of \ right \ classified \ clusters}{\# \ of \ total \ clusters}.
\end{equation}

The accuracy of the home location is 94.1\%, and that of the work location is 93.0\%. Further, by comparing the mobile phone results with survey data, we evaluated the accuracy of the inferred home locations at the aggregated level. Specifically, we collected population survey data for Beijing and Shanghai in 2015 (the same year of our mobile phone dataset). We then calculated the correlation between mobile phone inferred home locations and the population survey data at the district level. The results show that the $R^2$ values ($\log MobilePhone = \pi \log Survey + \epsilon$) are 0.97 for Beijing and 0.98 for Shanghai (Extended Data Fig. 7ab), implying strong consistency between the mobile phone estimated population and the real population distribution.

To verify whether the commuting distance derived from mobile phone data is consistent with that derived from other data sources, we collected Household Transportation Survey data in Shanghai from ref.\cite{zhang2016shanghai}. This survey was conducted in four areas of Shanghai (Jiading, Qingpu, Songjiang, and Jinshan) and covers a sample of 15,000 people ($\sim 1\%$ of the population in these areas). Zhang et al.\cite{zhang2016shanghai} computed the commuting patterns of Shanghai based on a mobile phone dataset spanning six months in 2013, and the results show that the mean commuting distance is 8.2 km, which is very close to our finding (Fig.~\ref{fig:1}b). Additionally, ref.\cite{zhang2016shanghai} provides a detailed comparison between mobile phone data and survey results in terms of commuting. As shown in Extended Data Fig. 7ac, the distributions of the distance between home and work locations derived by the two approaches are very similar. 

\subsection{Commuting distance and time within a city}\

We directly calculated the geodesic distance between home and work locations detected above, and we present the results in Fig.~\ref{fig:1}. However, different cities may have different road network structures and traffic conditions. Thus, we use online map services to obtain the real-time commute distance and time for the within city analysis. To simplify the calculation, we aggregate individual-level home-work data (denoted as origin-destination, OD for short) into $1km \times 1km$ grid cells and sum the \textit{commuting flows} between each OD pair to construct commuting networks. We remove OD pairs with a distance less than 1 km since they would not generate commuting flows under the $1km \times 1km$ grid cell setting. The commuting networks' geographical pattern is shown in Fig.~\ref{fig:1}a and Extended Data Fig. 3. For a given OD pair, we query the Baidu Map APIs (http://lbsyun.baidu.com/) for the real-time travel distance and time via two transportation modes: driving and public transit (e.g., bus or subway). Each OD pair is queried for one week of workdays starting from 7:30 am and ending before 9:30 am the same day. This process was conducted during June -- September 2018. We then take the mean and median values of the results to remove potential bias caused by extreme traffic situations, such as extreme weather or road accidents. There are 200k-300k requests for each large city. We set a high concurrency to complete all requests during the daily commuting period (7:30 am-9:30 am). Note that Baidu Map APIs take into account the real road conditions, and congestion is considered in the travel time. Thus, the average commuting distance and time calculated by this approach could be viewed as the typical commuting distance and duration of a given OD pair. For public transit, Baidu Map APIs take into account the transfer time and walking time. On the other hand, for the driving mode, parking and waiting time are not considered; thus, we add 10 minutes of parking and waiting time to the driving results, which does not change the shape of the curve but makes the total time closer to the real situation.  

Figure~\ref{fig:2} shows the within-city commuting results for Shanghai and Chengdu, and Extended Data Fig. 5 and Extended Data Fig. 6 show results for eight additional cities (Beijing, Guangzhou, Tianjin, Nanjing, Shenyang, Harbin, Hangzhou, and Changsha). The descriptive statistics of these cities are documented in Extended Data Table 1. Almost all cities have a maximum attractive radius of approximately 10 km from the city centre. However, Beijing appears to be an outlier, as its commuting distance peaks at a radius of 20 km. This may be partly due to the fact that Beijing has built a well-connected subway system that can support longer commutes. On the other hand, urban planning problems in Beijing (job-house balance issue) have resulted in a higher frequency of long-distance commuting.

\subsubsection*{Analytic results of the model} 
A second centre $s$ appears when, for an individual $j$, the utility of commuting to $s$ is higher than that of commuting to the first centre, $U_{js} > U_{j1}$, which leads to 

\begin{equation}
    1 - \beta^{*}d_{js} > E(1)^{\gamma} - \beta^{*}d_{j1}[1+ b(\frac{T(1)^{\eta}}{c})^\mu].
\label{u}
\end{equation}

Assuming there is no spatial correlation between each location, we have $d_{js} \approx d_{j1} \approx R(t) \sim N(t)^{1/3}$. We know that for the first centre, the population is equal to employment, $T(1) = E(t) = N(t) \gg 1$, indicating a critical value for the population in the stage of polycentric transition 

\begin{equation}
    N(t) \sim (\frac{c^\mu}{\beta^*})^{\frac{1}{\mu\eta - \gamma + 1/3}}.
\label{nt}
\end{equation}

Given Eq.~(\ref{nt}), there will always be a critical value of the population above which a city becomes polycentric. According to Eq.~(\ref{nt}), the attraction radius of the largest centre is derived as

\begin{equation}
    d \sim (\frac{c^\mu}{\beta^*})^{\frac{1/3}{\mu\eta - \gamma + 1/3}}.
\label{d}
\end{equation}

Eq.~(\ref{d}) shows that the most effective way to stretch the attraction radius of the city is to increase its transportation capacity $c$, for example, by building up more roads or adding a subway system. Such changes would shift the commuting curve to the right (the green line in Fig.~\ref{fig:4}f). Since $\beta^*$, $\mu$, $\eta$, and $\gamma$ are approximately constant across cities (The stablility $\beta^*$ is presented in the previous section, $\gamma$ is the scaling exponent between wage per capita and population size, and $\eta$  captures the relationship between road volume per capita and population size. In empirical data and theoretical models, $\gamma \approx \eta \approx 1/6$, see ref.\cite{li2017simple,prud1999size,bettencourt2013origins}.), the attraction radius $d$ then is almost constant. As shown in Fig.~\ref{fig:2}, Extended Data Fig. 5 and Extended Data Fig. 6, the radius of the main centre in these Chinese cities is approximately 10 km, with a relatively larger value for cities with strong subway systems, such as Bejing and Shanghai. Again, this empirical result is in full accordance with Eq.~(\ref{d}). Furthermore, Eq.~(\ref{d}) indicates that if the agglomeration effect $\gamma$ of a city is relatively large, that cities should have a larger attraction radius.

In Fig.~\ref{fig:1}b, as the population increases in cities with a subway system (red), the commuting distance tends to increase ($p$-value $<$ 0.05). Under the constant travel/commuting time hypnosis and according to several empirical findings\cite{levtnson1997density, anas2015urban}, taking the subway can avoid the congestion of the road traffic and therefore extend the commute distance for a given commute time. 

\clearpage

\begin{figure*}[!ht]
    \centering
    \includegraphics[width=.85\linewidth]{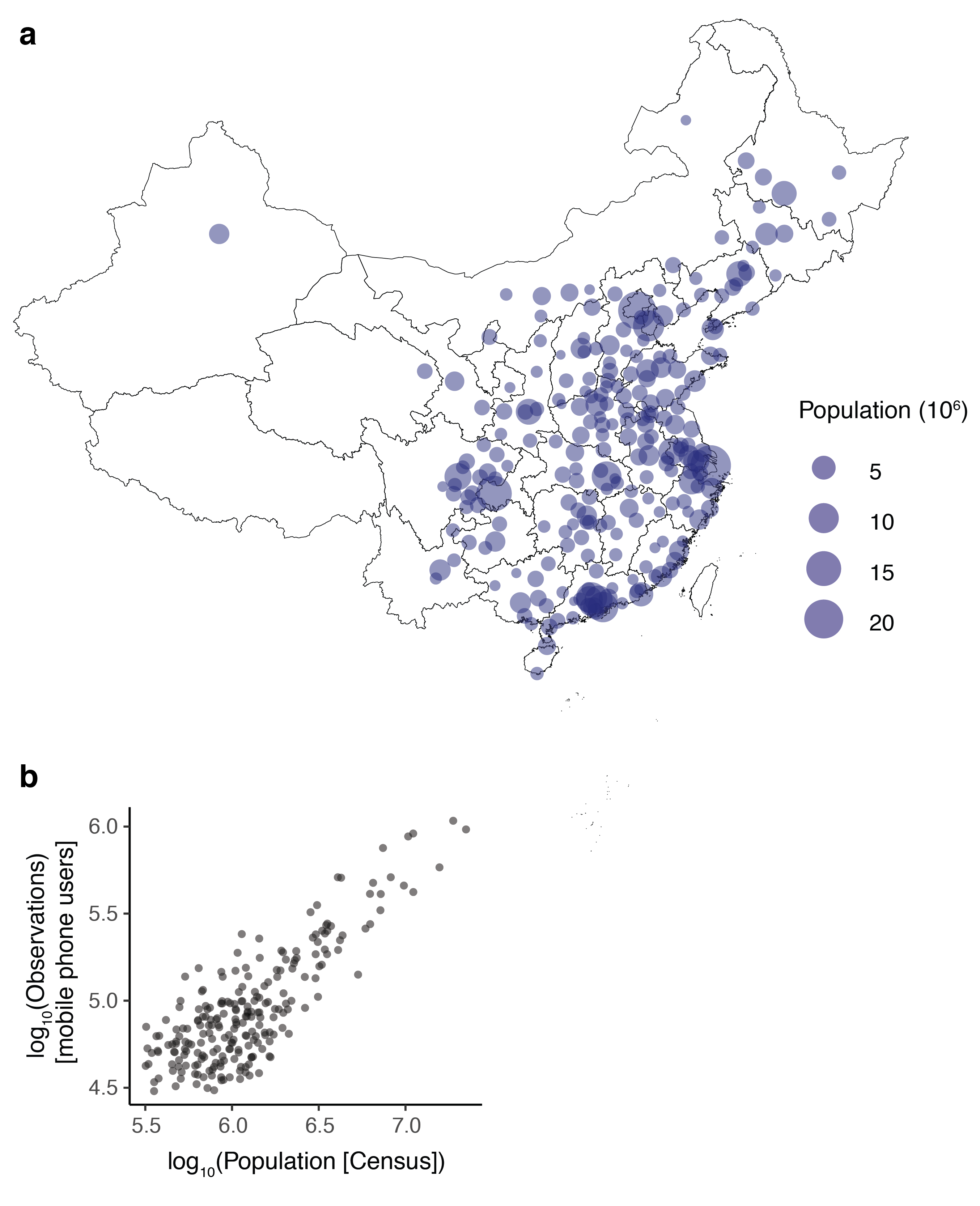}
    \begin{flushleft}
    \textbf{Extended Data Fig. 1 Studied cities.} \textbf{a,} Spatial distribution of the studied prefecture-level cities. Node sizes represent the 2010 Census urban population. These cities are located in different geographic regions and vary greatly in population size. \textbf{b,} Relationship between census urban population and mobile phone users in our dataset. The Pearson correlation between these two variables is $r = 0.903$ (n = 234). 
    \end{flushleft}
    \label{fig:city}
\end{figure*}

\clearpage
\begin{figure*}[!ht]
    \centering
    \includegraphics[width=1.\linewidth]{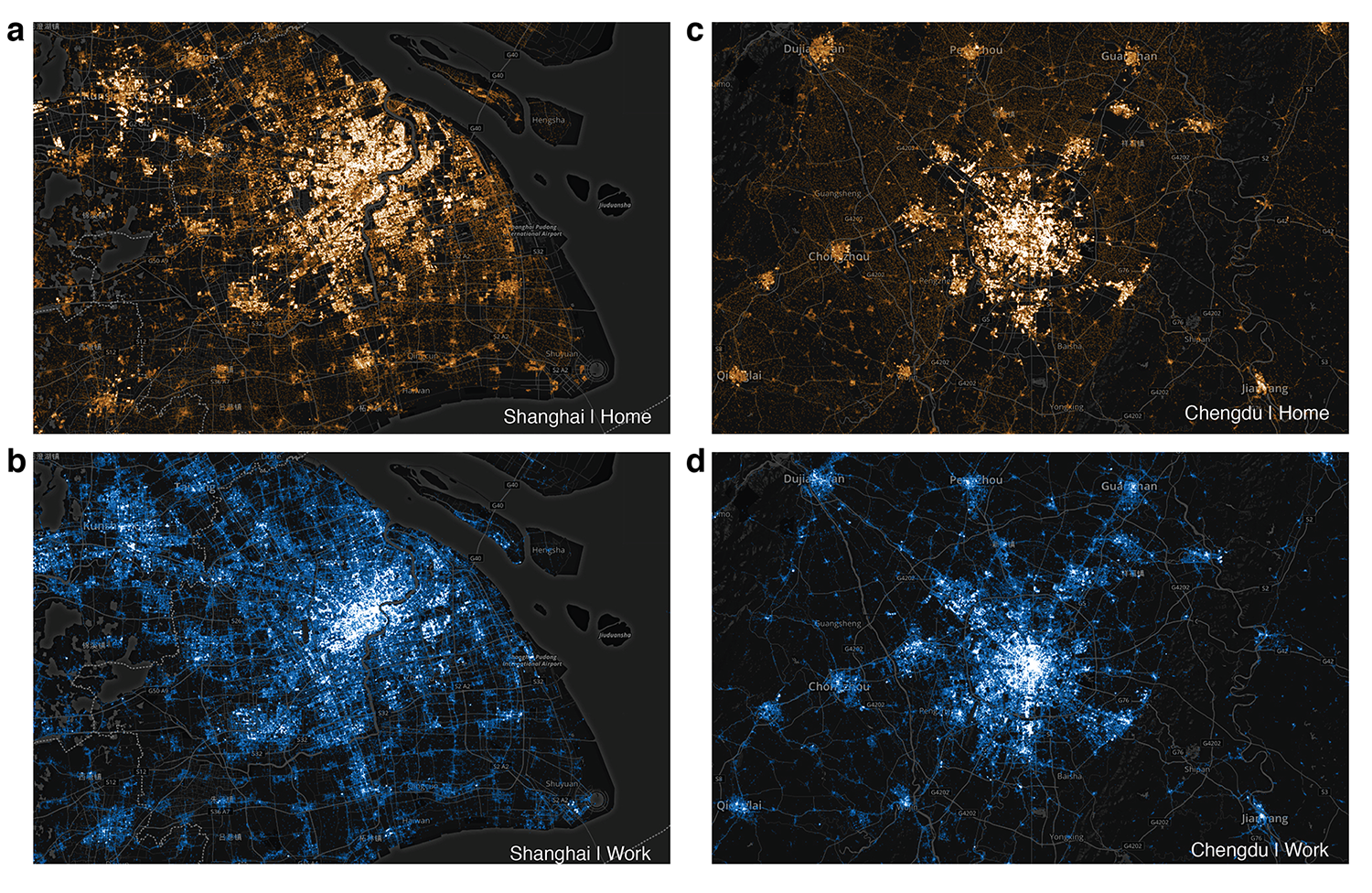}
    \begin{flushleft}
    \textbf{Extended Data Fig. 2 Spatial distribution of home and work locations.} Home (\textbf{a, c}) and work (\textbf{b, d}) locations for Shanghai and Chengdu. The work (population) density around the city centre is much higher than the home (population) density due to agglomeration. 
    \end{flushleft}
    \label{fig:home_work}
\end{figure*}

\clearpage
\begin{figure*}[!ht]
    \centering
    \includegraphics[width=1.\linewidth]{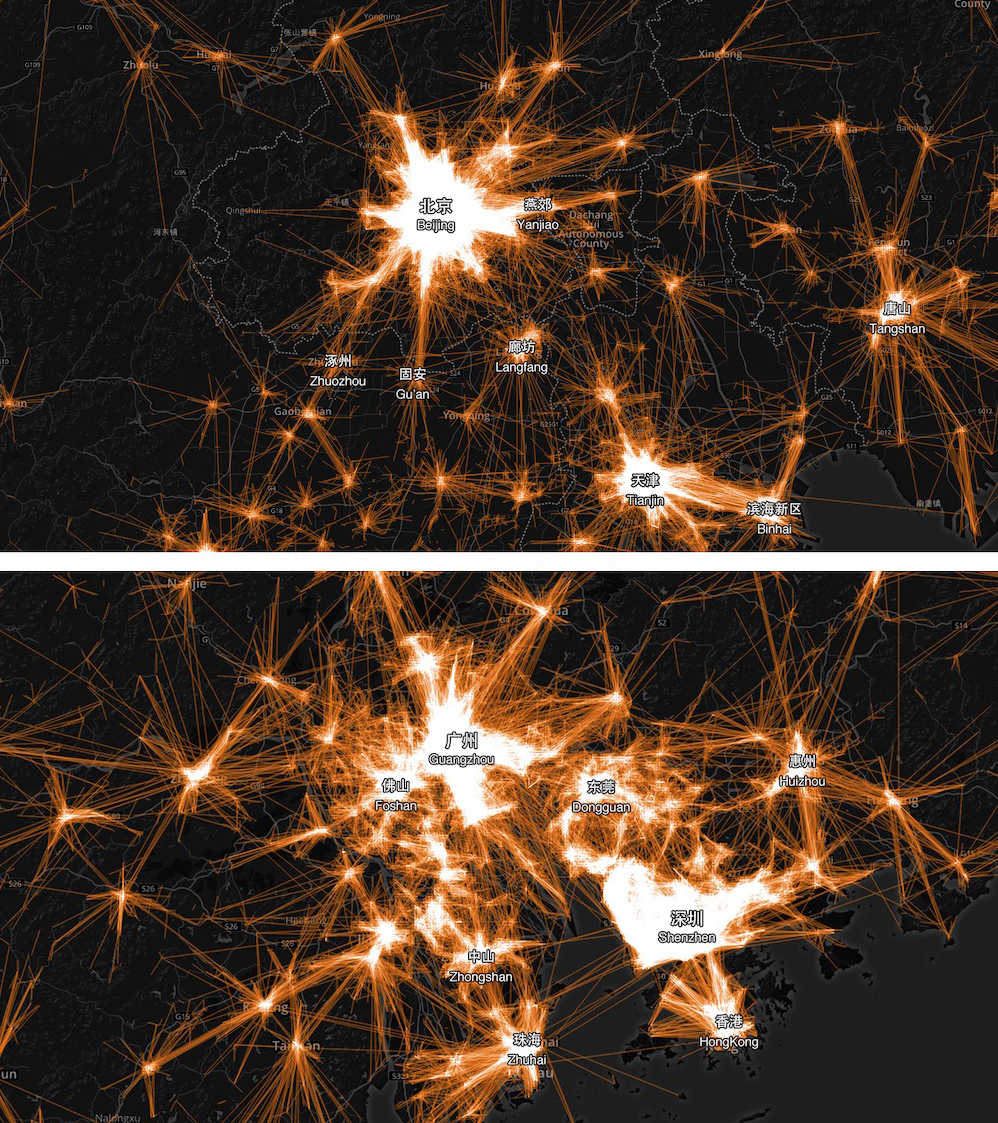}
    \begin{flushleft}
    \textbf{Extended Data Fig. 3 Commuting networks.} Beijing-Tianjin-Hebei area (upper) and Pearl River Delta (lower).
    \end{flushleft}
    \label{fig:commute_network}
\end{figure*}

\clearpage
\begin{figure*}[!ht]
    \centering
    \includegraphics[width=1.\linewidth]{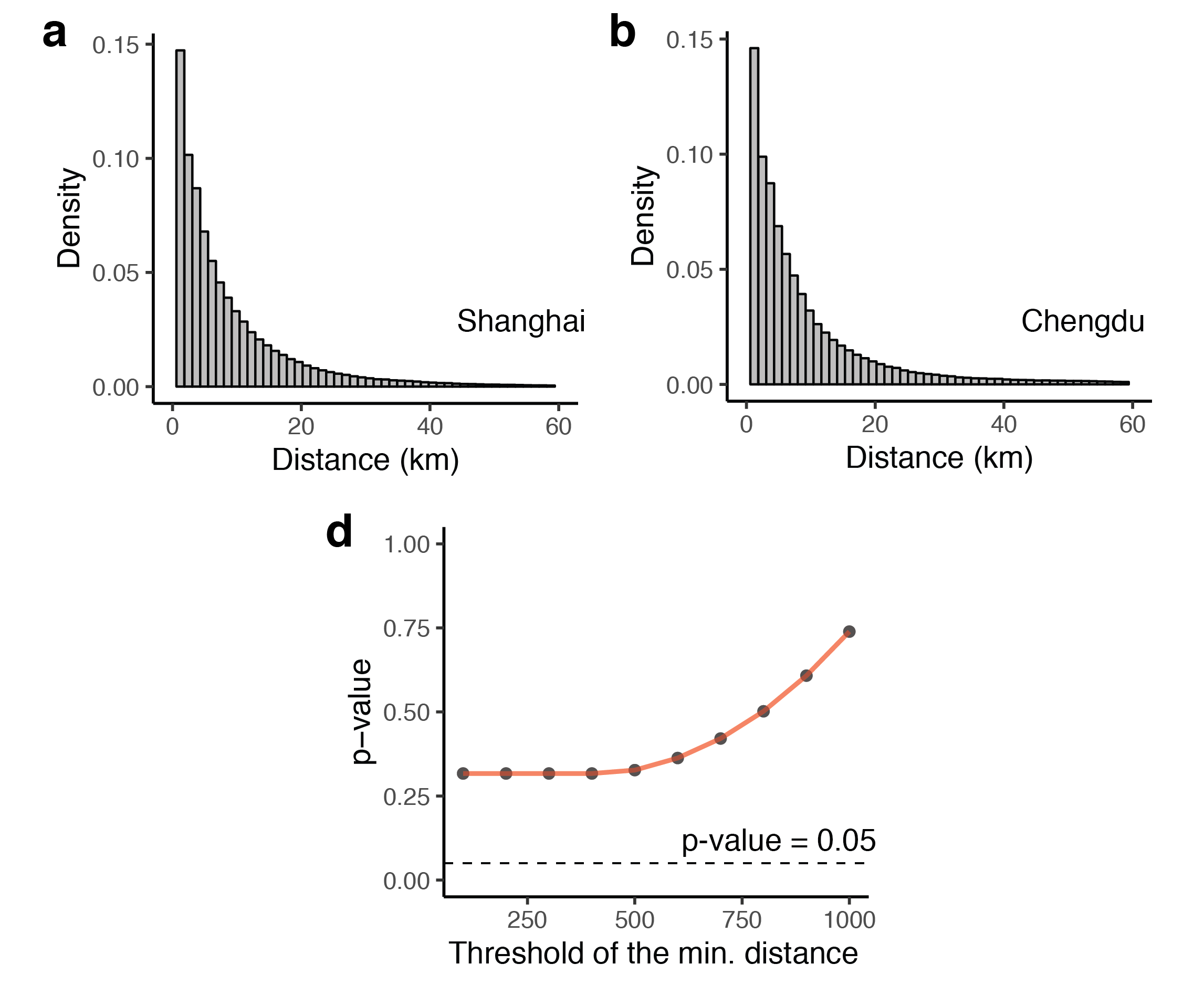}
    \begin{flushleft}
    \textbf{Extended Data Fig. 4 Commuting distance distribution and robustness check.} Commuting distance histogram plots of Shanghai (\textbf{a}) and Chengdu (\textbf{b}). \textbf{c,} Robustness check of the scale-invariant commuting distance. We drop users with commuting distance below a threshold (from 100 m to 1000 m), and calculate the p-value of the Person correlation between the city average commuting distance and city population size. All p-values are $> 0.3$, indicating the two variables are uncorrelated.
    \end{flushleft}
    \label{fig:robustcheck}
\end{figure*}

\clearpage
\begin{figure*}[!ht]
    \centering
    \includegraphics[width=1.\linewidth]{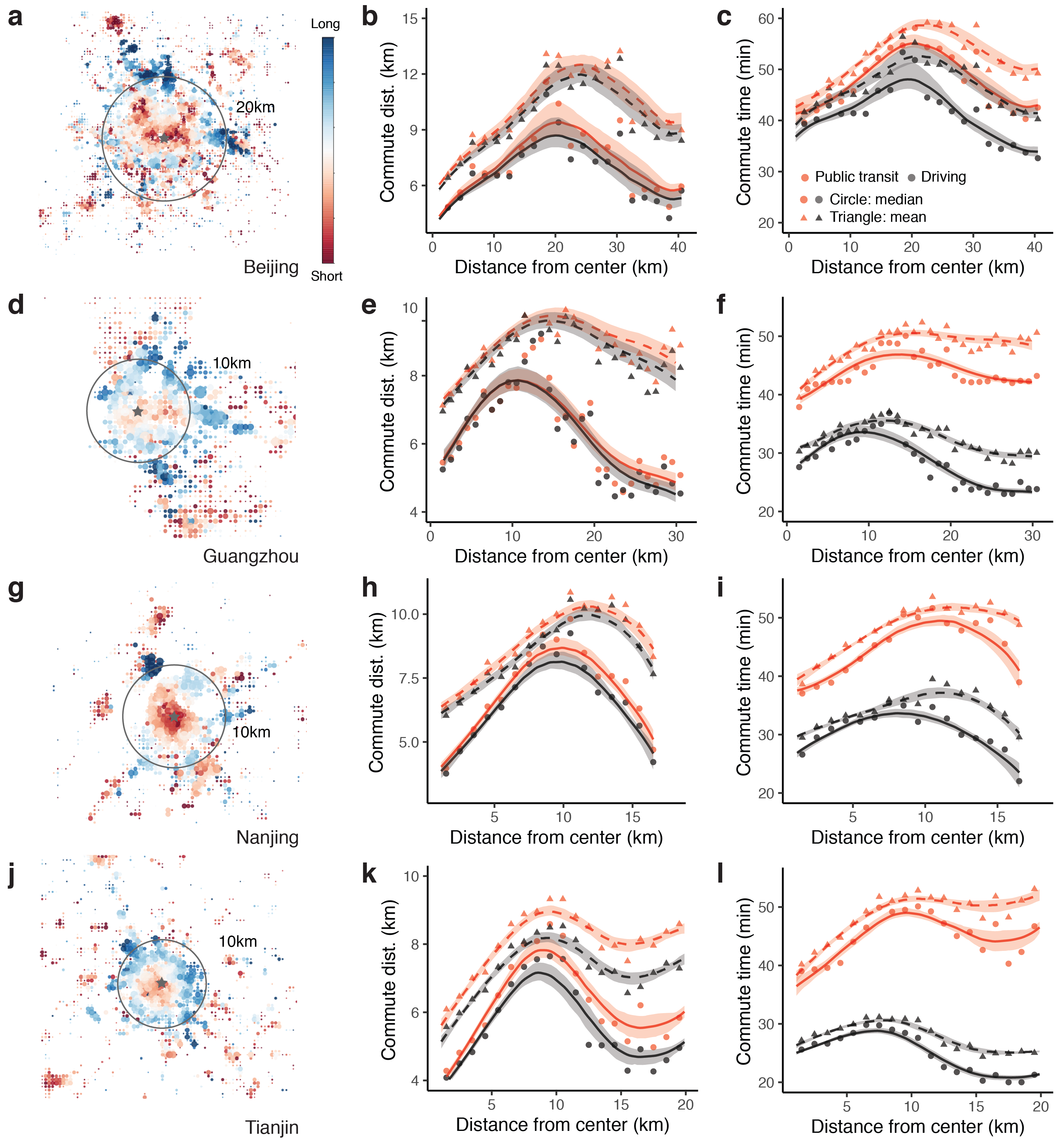}
    \begin{flushleft}
    \textbf{Extended Data Fig. 5 Commuting patterns within cities (1)}. Spatial distribution (by home location) of commuting distance in Beijing (\textbf{a}), Guangzhou (\textbf{d}), Nanjing (\textbf{g}), and Tianjin (\textbf{j}). City centres are marked with star symbols. Average commuting distance/time (y-axis) with varying distance from the city centre (x-axis) of Beijing (\textbf{b, c}), Guangzhou (\textbf{e, f}), Nanjing (\textbf{h, i}), and Tianjin (\textbf{k, l}). Red colour represents commuting by public transit, black is by driving, circle and triangle symbols represent mean and median results, respectively. We also apply non-parametric fitting for each scatter plot with the \textit{npregfast} library in R and calculate the 95\% confidence interval (shaded areas) of each fitting curve by the bootstrapping method.
    \end{flushleft}
    \label{fig:ushape1}
\end{figure*}

\clearpage
\begin{figure*}[!ht]
    \centering
    \includegraphics[width=1.\linewidth]{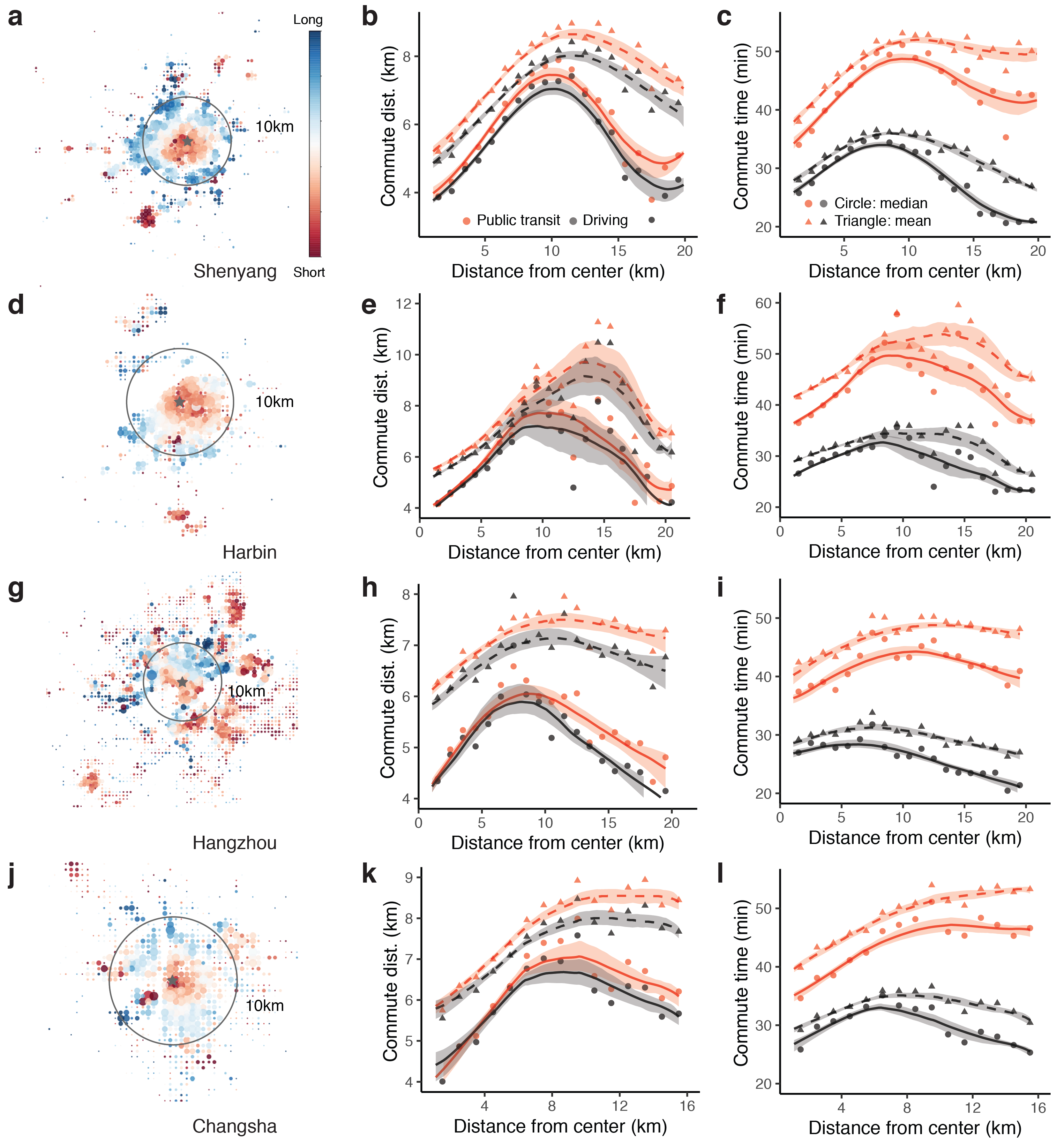}
    \begin{flushleft}
    \textbf{Extended Data Fig. 6 Commuting patterns within cities (2)}. Similar to Fig. S4. Shenyang (\textbf{a-c}), Harbin (\textbf{d-f}), Hangzhou (\textbf{g-i}), and Changsha (\textbf{j-l}).
    \end{flushleft}
    \label{fig:ushape2}
\end{figure*}

\clearpage
\begin{figure*}[!ht]
    \centering
    \includegraphics[width=1.\linewidth]{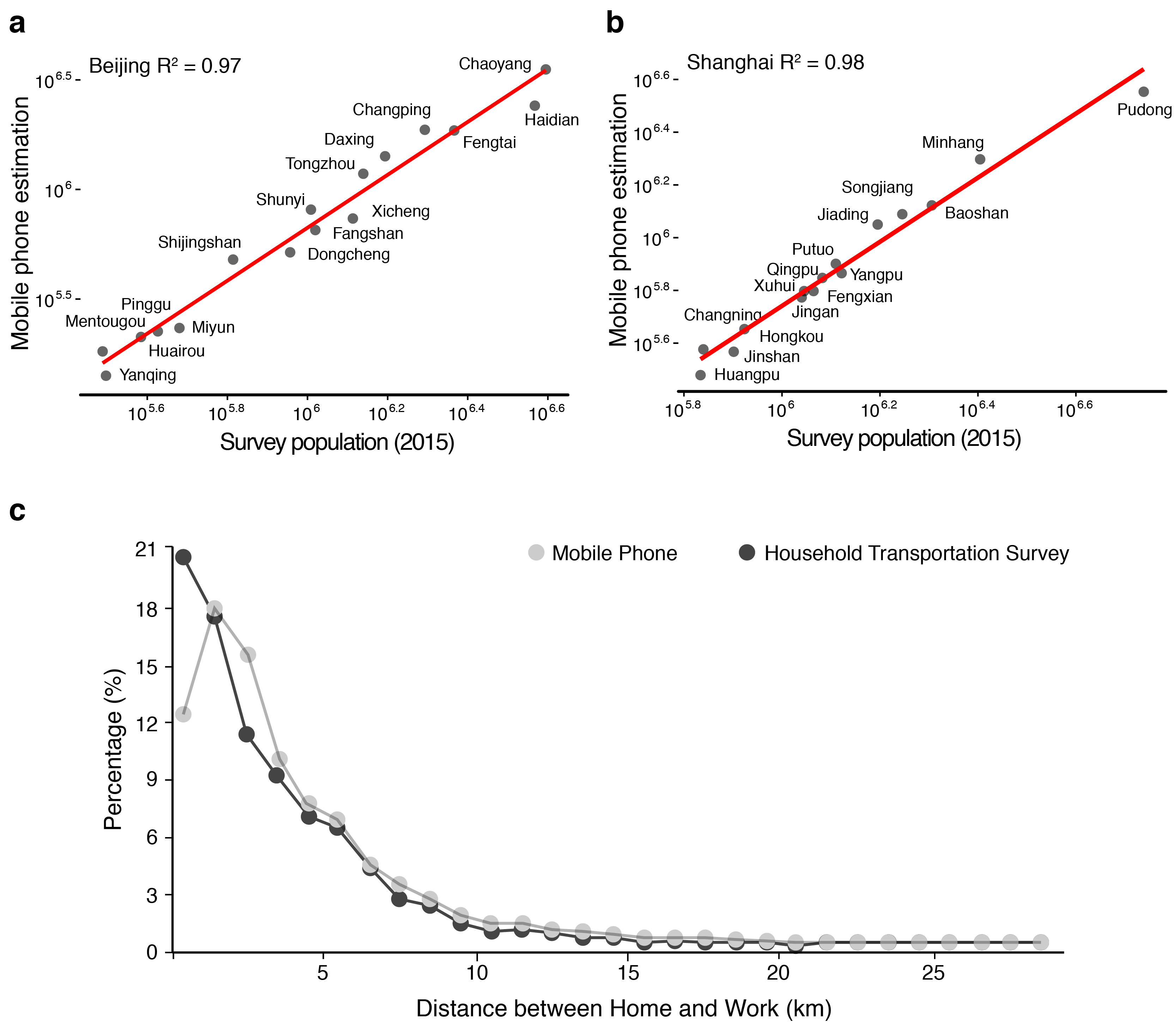}
    \begin{flushleft}
    \textbf{Extended Data Fig. 7 Data validation.} \textbf{a, b,} The population survey and mobile phone estimated population for Beijing (\textbf{a}) and Shanghai (\textbf{b}). \textbf{c,} Transportation survey and mobile phone inferred commuting distances (data source: ref \cite{zhang2016shanghai}).
    \end{flushleft}
    \label{fig:validation}
\end{figure*}

\clearpage
\begin{figure*}[!ht]
    \centering
    \includegraphics[width=.6\linewidth]{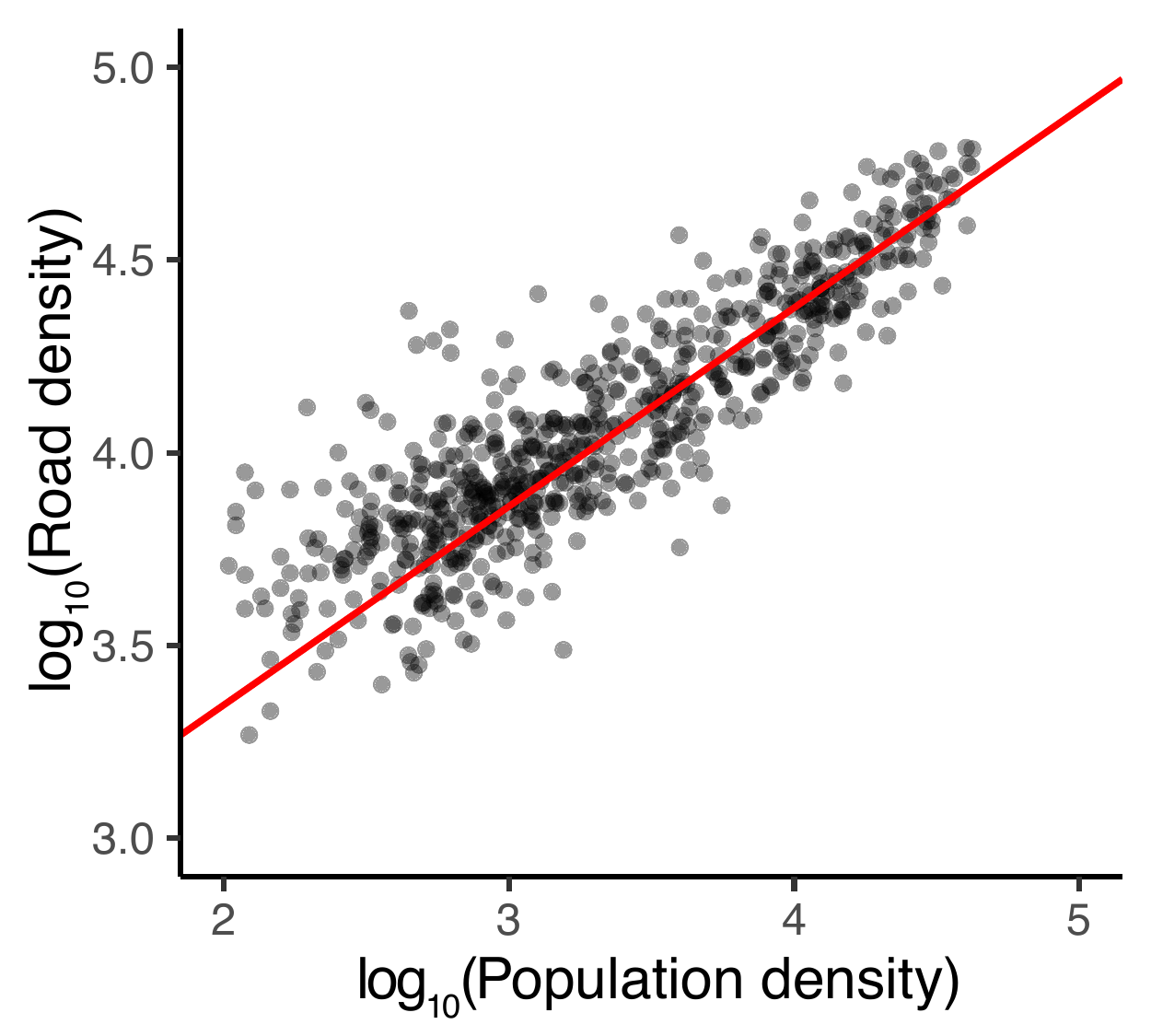}
    \begin{flushleft}
    \textbf{Extended Data Fig. 8 Scaling relationship between population density and road density (Chengdu).} Grid size is $2km \times 2km$, slope = 0.515 (0.020), and $r^2 = 0.689$.
    \end{flushleft}
    \label{fig:s9}
\end{figure*}

\clearpage

\begin{table*}[!ht]
    \footnotesize
    \centering
    \begin{tabular}{l|l}
    \hline
        Category & Features \\
    \hline
    Individual level & Number of stay points \\
         & Number of unique dates of stay points \\
         & Weekday number of stay points / weekend number of stay points \\
         & Weekday daytime number of  stay points / weekday nighttime number of stay points \\
    \hline
    Cluster level & Weekday number of stay points / Weekend number of stay points (each cluster) \\
         & Weekday daytime number of stay point / weekday nighttime number of stay points (each cluster)\\
         & number of stay points in each cluster / total number of stay points \\
         & Weekday number of stay points (each cluster) / total number of stay points \\
         & Weekend number of stay points (each cluster) / total number of stay points \\
         & Daytime number of stay points (each cluster) / total number of stay points \\
         & Nighttime number of stay points (each cluster) / total number of stay points \\
         & Number of other clusters to this cluster before 12:00 (transfer matrix)\\
         & Number of this cluster to other clusters before 12:00 (transfer matrix)\\
         & Number of other clusters to this cluster after 12:00 (transfer matrix)\\
         & Number of this cluster to other clusters after 12:00 (transfer matrix)\\
    \hline
    POI level & Number of residential points of interest (POIs)\\
         & Number of working points of interest (POIs)\\
    \hline
    \end{tabular}
    \begin{flushleft}
    \normalsize{\textbf{Extended Data Table 1 Main features for home and work location classification.} We set 9:00-18:00 as daytime, and the remaining period as nighttime; Monday-Friday as weekday, and Saturday and Sunday are weekend. Note that `transfer matrix' at cluster level means movement between clusters.}
    \end{flushleft}
    \label{tab:s4}
\end{table*}

\clearpage

\begin{table*}[!ht]
    \centering
    \begin{tabular}{c|c|c|c|c|c}
    \hline
        City & City centre & Latitude & Longitude & Urban population & Built-up area \\
         & & & & ($\times 10^6$)& (km$^2$) \\
    \hline
        Beijing & Tian'anmen Square & 39.907 & 116.391 & 18.82 & 2598 \\
        Changsha & Furong Square & 28.192 & 112.971 & 3.09 & 269 \\
        Chengdu & Tianfu Square & 30.659 & 104.064 & 7.42 & 690\\
        Guangzhou & People Park & 23.129 & 113.259 & 11.07 & 1144 \\
        Hangzhou & Wulin Square & 30.273 & 120.159 & 6.24 & 969\\
        Harbin & Train Station & 45.758 & 126.625 & 5.88 & 657\\
        Nanjing & Xinjiekou & 32.043 & 118.779 & 7.17 & 560 \\
        Shanghai & People's Square & 31.231 & 121.471 & 22.32 & 2174 \\
        Shenyang  & Shenyang Square & 41.797 & 123.428 & 6.26 & 794 \\
        Tianjin & Heping Road & 39.124 & 117.198 & 11.09 & 1821\\
    \hline
    \end{tabular}
    \begin{flushleft}
    \textbf{Extended Data Table 2 Descriptive statistics of ten cities.} Coordinates of city centres are derived from Wikipedia, Google Earth, and OpenStreetMap. Urban population data are from the 2010 Census. Built-up area data are from the World Bank PUMA database (https://puma.worldbank.org/). 
    \end{flushleft}
    \label{tab:s1}
\end{table*}

\clearpage

\end{document}